\documentclass{article}
\usepackage[preprint]{spconf}
\usepackage{amsmath,graphicx}
\usepackage{amsfonts}
\usepackage[normalem]{ulem}

\usepackage{xcolor}
\usepackage{diagbox}
\usepackage{adjustbox}
\usepackage{multirow}
\usepackage{booktabs}
\usepackage{mathtools}
\usepackage{amsmath}
\usepackage{float}
\usepackage{cite}
\usepackage{subcaption}
\usepackage{hyperref}
\usepackage{bbm}
\usepackage{algorithm}
\usepackage{algorithmic}
\usepackage[marginal]{footmisc}
\usepackage{bm}

\usepackage[normalem]{ulem}
\useunder{\uline}{\ul}{}

\newcommand{\mycomment}[1]{}

\title{Revealing Emotional Clusters in Speaker Embeddings: \\ A Contrastive Learning Strategy for Speech Emotion Recognition \vspace{-0.2cm}}

\name{Ismail Rasim Ulgen$^1$, Zongyang Du$^1$, Carlos Busso$^2$, Berrak Sisman$^1$ \vspace{-0.2cm}} 

\address{$^1$Speech \& Machine Learning (SML) Lab, The University of Texas at Dallas, USA \\
$^2$Multimodal Signal Processing (MSP) Lab, The University of Texas at Dallas, USA
\vspace{-0.2cm}}

\begin{document}
%

\copyrightnotice{\  {\parbox{\dimexpr\textwidth-\fboxsep-\fboxrule\relax}{\scriptsize \copyright  2024 IEEE. Personal use of this material is permitted. Permission from IEEE must be obtained for all other uses, in any current or future media, including reprinting/republishing
this material for advertising or promotional purposes, creating new collective works, for resale or redistribution to servers or lists, or reuse of any copyrighted component of this work
in other works.}}}
\maketitle

\begin{abstract}
\vspace{-1mm}
Speaker embeddings carry valuable emotion-related information, which makes them a promising resource for enhancing speech emotion recognition (SER), especially with limited labeled data. Traditionally, it has been assumed that emotion information is indirectly embedded within speaker embeddings, leading to their under-utilization. Our study reveals a direct and useful link between emotion and state-of-the-art speaker embeddings in the form of intra-speaker clusters. By conducting a thorough clustering analysis, we demonstrate that emotion information can be readily extracted from speaker embeddings. In order to leverage this information, we introduce a novel contrastive pretraining approach applied to emotion-unlabeled data for speech emotion recognition. The proposed approach involves the sampling of positive and the negative examples based on the intra-speaker clusters of speaker embeddings. The proposed strategy, which leverages extensive emotion-unlabeled data, leads to a significant improvement in SER performance,  whether employed as a standalone pretraining task or integrated into a multi-task pretraining setting.


\end{abstract}
\vspace{-2mm}
\vspace{-4mm}
\section{Introduction}
\label{sec:intro}
\vspace{-3mm}
\mycomment{
\textbf{Paragraph 1:} \\
\textit{SER definition \\
SER challenging conditions }\\}
Speech emotion recognition remains a challenging task due to its complexity and the subjective nature of emotional expression, compounded by the scarcity of labeled emotional data \cite{Sethu2019TheAW}. These factors significantly hinder the development of effective SER methods, and encourage researchers to leverage auxiliary knowledge from closely related speech tasks, such as speaker verification (SV) \cite{Pappagari2020XVectorsME, Padi2021ImprovedSE, Zhou2020ATL, pepino21_interspeech}.

\par
\mycomment{
\textbf{Paragraph 2:} \\
\textit{Speaker verification definition \\
Speaker verification does not have those problems\\}
common objectives with SER \\
Speaker Embeddings are utilized in SER \\}


In contrast to SER, SV benefits from the availability of sufficient labeled data \cite{Nagrani2017VoxCelebAL, Chung2018VoxCeleb2DS}.  Although the tasks of recognizing emotions from speech and verifying speakers differ in their primary objectives, they both revolve around the identification of fundamental voice attributes, including pitch, tone, and phonation patterns. Consequently, speaker verification techniques with robust performance are now being explored as promising tools for enhancing the performance of speech emotion recognition systems \cite{Pappagari2020XVectorsME,Padi2021ImprovedSE,Aldeneh2023YoureNY}\mycomment{ \textbf{Is this what we believe, or are there citations? Do we want to use this as motivation to our paper?}.}
 
\mycomment{
\textbf{Paragraph 3:} \\
\textit{Analysis of emotion in speaker embeddings \\
Utilization of SV in SER (details)}\\}
Emotion information within speaker features has been explored in various emotional speech tasks. Studies \cite{Parthasarathy2017ASO,Parthasarathy2017PredictingSR, Bancroft2019ExploringTI} revealed increased equal error rates in speaker verification for non-matching emotional conditions, highlighting the sensitivity of speaker features to emotional states \cite{Dehak2011FrontEndFA}. Research by \cite{Williams2019DisentanglingSF} demonstrated emotion-related information in speaker embeddings via autoencoder-based reconstruction analysis and emotion classification. This finding was confirmed by \cite{Aldeneh2023YoureNY}, which also performed reconstruction analysis and used speaker embeddings as SER input features. Recent works \cite{Pappagari2020XVectorsME, Padi2021ImprovedSE} employed deep speaker embedding networks to transfer knowledge from speaker verification to speech emotion recognition. However, the potential of recent deep speaker embeddings like d-vector \cite{Wan2017GeneralizedEL} and ECAPA-TDNN \cite{desplanques20_interspeech} in encoding emotional information remains an area that requires comprehensive exploration. Previous studies are limited by the assumption that emotion information is indirectly encoded within speaker embeddings and can be utilized under supervision. In this paper, we aim to explore whether emotion-related information directly resides within the speaker embedding space and find effective ways to leverage this information in SER tasks.

\begin{table*}[!ht]
\vspace{-4mm}
\caption{Intra-speaker clustering evaluation for emotion classification.} \vspace{-4mm}
\label{tab:clustering}
\centering
\scalebox{0.8}{\begin{tabular}{c|cccc|cccc}
\hline
{\ul }           & \multicolumn{4}{c|}{\textbf{d-vector}}                                                                                                                                     & \multicolumn{4}{c}{\textbf{ECAPA-TDNN}}                                                                                                                                  \\ \hline
\textbf{Dataset} & \multicolumn{1}{c|}{\textbf{NMI [0,1] $\uparrow$}} & \multicolumn{1}{c|}{\textbf{ARI [0,1] $\uparrow$}} & \multicolumn{1}{c|}{\textbf{Purity [0,1] $\uparrow$}} & \textbf{Silhoutte [-1,1] $\uparrow$} & \multicolumn{1}{c|}{\textbf{NMI [0,1] $\uparrow$}} & \multicolumn{1}{c|}{\textbf{ARI [0,1] $\uparrow$}} & \multicolumn{1}{c|}{\textbf{Purity [0,1] $\uparrow$}} & \textbf{Silhoutte [-1,1] $\uparrow$} \\ \hline
ESD        & \multicolumn{1}{c|}{0.76}                    & \multicolumn{1}{c|}{0.72}                    & \multicolumn{1}{c|}{0.89}                     & 0.14                         & \multicolumn{1}{c|}{0.89}                   & \multicolumn{1}{c|}{0.91}                   & \multicolumn{1}{c|}{0.97}                     & 0.13                         \\
IEMOCAP   & \multicolumn{1}{c|}{0.29}                    & \multicolumn{1}{c|}{0.21}                    & \multicolumn{1}{c|}{0.66}                     & 0.01                        & \multicolumn{1}{c|}{0.31}                   & \multicolumn{1}{c|}{0.25}                   & \multicolumn{1}{c|}{0.67}                     & 0.01                        \\
CREMA-D   & \multicolumn{1}{c|}{0.43}                    & \multicolumn{1}{c|}{0.39}                    & \multicolumn{1}{c|}{0.63}                     & 0.07                         & \multicolumn{1}{c|}{0.36}                   & \multicolumn{1}{c|}{0.27}                   & \multicolumn{1}{c|}{0.57}                     & 0.04                         \\
RAVDESS   & \multicolumn{1}{c|}{0.59}                    & \multicolumn{1}{c|}{0.38}                    & \multicolumn{1}{c|}{0.67}                     & 0.14                         & \multicolumn{1}{c|}{0.51}                   & \multicolumn{1}{c|}{0.28}                   & \multicolumn{1}{c|}{0.62}                     & 0.05                         \\ \hline

\end{tabular}}
\vspace{-5mm}
\end{table*}

\mycomment{
\textbf{A strong SER paraghraph,recent approaches for speech emotion recognition, unlabeled emotion data, and a transition to contrastive learning... End with "in this paper, we study contrastive learning for unlabeled...."}}

Self-supervised speech models such as wav2vec2.0 \cite{wav2vec2} can leverage large unlabeled speech datasets to enhance supervised SER frameworks \cite{pepino21_interspeech, Morais2022SpeechER, Li2021ContrastiveUL}. However, it's important to note that these pre-training objectives were not originally designed for SER, except for \cite{goncalves22_interspeech} which incorporated audiovisual features. Additionally, existing pretraining tasks utilized in SER are frame-level tasks while speech emotion is usually formulated as an uttterance-level task. Consequently, a significant gap exists in the field, particularly in the development of an utterance-level, unsupervised pre-training strategy explicitly tailored to SER, exclusively using speech-related features, which is one of the contributions of this paper.

This paper marks the first attempt to investigate the direct accessibility of emotion-related information within state-of-the-art deep speaker embeddings. Our analysis reveals distinct intra-speaker clusters that reflect emotional states, suggesting a strong link between speaker and emotion recognition. To utilize this information, we propose a novel pretraining strategy using large-scale, emotion-unlabeled data. This approach employs contrastive learning, forming positive-negative pairs based on speaker embedding clusters, without the need for emotion labels. We apply this strategy both as the primary objective of pretraining and as an additional task for the existing pretraining methods in a multi-task setting.  Our contributions can be summarized as follows: 1) We reveal readily available emotion information within speaker embeddings; 2) We introduce a unique, utterance-level contrastive learning approach for SER, without relying on emotion labels; 3) We demonstrate that combination of pretraining tasks in a multi-task setting can further improve SER performance; and 4) Through our proposed training strategy, we enhance a very strong framework, wav2vec2.0, in terms of emotion recognition performance.

\vspace{-4mm}
\section{Revealing Emotion Clusters in Speaker Embeddings}
\vspace{-2mm}

In this section, we conduct clustering analysis on speaker embeddings to explore emotion discrimination within the speaker embedding space, aiming to establish a direct link between intra-speaker clusters of embeddings and emotional categories. This connection holds significant potential for various SER applications, particularly in harnessing extensive, emotion-unlabeled data. Our analysis is driven by the hypothesis that speaker embeddings, designed to capture voice characteristics, are sensitive to variations in a speaker's voice across different emotional states\cite{Parthasarathy2017ASO,Parthasarathy2017PredictingSR, Aldeneh2023YoureNY, Williams2019DisentanglingSF}, drawing inspiration from studies indicating distinct speaker patterns in different emotional contexts\cite{Parthasarathy2017ASO,Parthasarathy2017PredictingSR, Bancroft2019ExploringTI}. 

\vspace{-4mm}
\subsection{Dataset, speaker embeddings and evaluation metrics}
\vspace{-1mm}
\label{clus_exp}

We applied k-means clustering to length-normalized speaker embeddings of using a maximum of 320 different utterances for each speaker within a dataset, using a fixed number of clusters to align with the four categorical emotions: neutral, happiness, sadness, and anger. We selected four widely used labeled emotion datasets: IEMOCAP\cite{busso2008iemocap}, ESD\cite{zhou2021emotional}, CREMA-D\cite{6849440}, and RAVDESS\cite{livingstone2018ryerson}. Our choice of deep speaker embedding networks includes d-vector\cite{Wan2017GeneralizedEL} and ECAPA-TDNN\cite{desplanques20_interspeech}, both trained with metric-based objectives like generalized end-to-end loss and angular margin softmax loss on the voxceleb2 dataset\cite{Chung2018VoxCeleb2DS}. We evaluated the alignment between intra-speaker cluster labels and emotion categories using metrics such as Normalized Mutual Information (NMI)\cite{Nguyen2010InformationTM}, Adjusted Rand Index (ARI)\cite{Nguyen2010InformationTM}, Purity Score\cite{rendon_purity11}, and Silhouette Score\cite{Rousseeuw1987SilhouettesAG}, averaged over speakers and larger values indicate a stronger alignment. 

\vspace{-4mm}
\subsection{Clustering and Evaluations}
\vspace{-2mm}
\mycomment{\textbf{Can you explain what we are observing here and why we do these experiments}}

The clustering results are reported in Table \ref{tab:clustering}. Notably, the ESD dataset consistently demonstrates exceptionally high metrics, indicating a direct alignment between intra-speaker clusters and emotion categories in specific conditions where the utterances are very clean, linguistic content is normalized over emotion categories and emotion intensity tends to be high. While the metrics for other datasets are not as high as in ESD, a meaningful correlation exists across all datasets. The IEMOCAP dataset, with challenges like reverberation and overlapping speech, exhibits the lowest metrics, possibly due to variance introduced into speaker embeddings.

\begin{figure}[tbp]
    \centering
    \begin{subfigure}{0.5\textwidth}
     \centering
    \includegraphics[width=5.7cm]{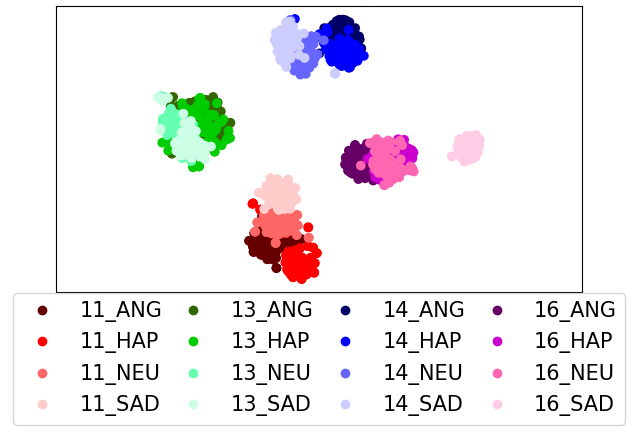}
    \vspace{-2mm}
    \caption{T-SNE of speaker embeddings in ESD dataset}
    \label{esd}
    \end{subfigure}    \vspace{-3mm}
    
    \begin{subfigure}{0.5\textwidth}
     \centering
    \includegraphics[width=5.7cm]{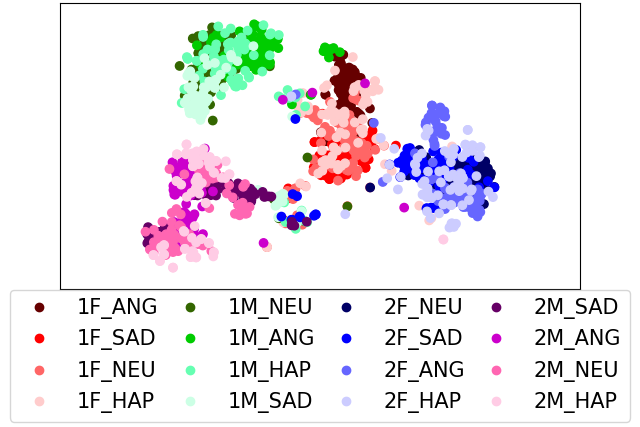}
    \vspace{-2mm}
    \caption{{T-SNE of speaker embeddings in IEMOCAP dataset}}
    \label{iemocap}
    \end{subfigure}  
    \vspace{-6mm}
    \caption{Visualization of intra-speaker clusters in two datasets, the colors represent \{speaker id\}\_\{emotion\}.}
    \vspace{-6mm}
    \label{fig:spk_emb}
\end{figure}

The distribution of embeddings can be observed in the t-SNE plots in Figure \ref{fig:spk_emb}, showing clear separation in the ESD dataset and some distinction in the IEMOCAP dataset. We've plotted t-SNE plots only for ESD and IEMOCAP due to similar trends in other databases. NMI values tend to be higher than ARI values, indicating uneven clustering errors. Higher purity values, compared to lower ARI values, suggest overlaps between specific emotion pairs, hinting at unique relationships between emotion categories. Low silhouette scores are expected due to closely spaced embeddings, aligning with their original goal of grouping speaker utterances together.

In general, the clustering results validate that speaker embeddings tend to group together for different emotional states in the embedding space due to distinct vocal characteristics for each emotion. The correspondence between emotion categories and intra-speaker clusters is limited in non-ideal conditions possibly due to other factors affecting the speech signal.\mycomment{MOVE TO CONCLUSION: The effect of other factors such as noise, and linguistic content deserves a detailed analysis which could be very enlightening.} 
The results show that even clusters with limited accuracy can serve as effective learning tasks \cite{hsu2021hubert, Han2023SelfSupervisedLW, Tao2021SelfSupervisedSR}. Inspired by these findings, we propose a contrastive learning strategy based on the trend of intra-speaker clustering of emotion categories.
\vspace{-2mm}

\label{sec:pagestyle}




\vspace{-3mm}
\section{Contrastive Learning for SER}
\vspace{-3mm}
\mycomment{\textbf{You should start by talking about the motivation - a transition from section 2 here}}
\mycomment{\textcolor{orange}{transition from section 2 to section 3. }}
In this study, we introduce a novel contrastive pretraining strategy without emotion labels, which capitalizes on emotion-related information present in the form of intra-speaker clusters within speaker embeddings. Our approach is based on contrastive learning, a technique well-known for its efficacy across various tasks \cite{simclr, He2019MomentumCF}. The learning objective tries to maximize the similarity between positive pairs while minimize it for negative pairs. In our approach, positive pairs consist of utterances sampled from the same intra-speaker cluster, likely sharing the same emotion category. In contrast, negative examples are created from different
intra-speaker clusters of the same speaker, likely to have different emotion categories given our analysis in Section 2. This setup inherently fosters an utterance-level emotion classification.


\begin{figure}[t]
    \centering
    \scalebox{0.61}{\includegraphics{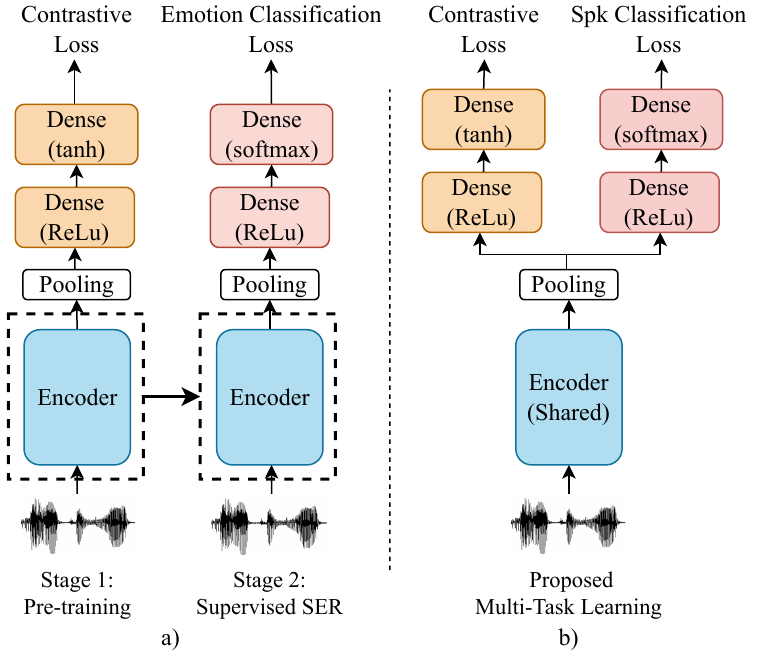}}
    \vspace{-4mm}
    \caption{a) Proposed contrastive pre-training and SER training, b) Proposed multi-task learning framework.}
    \vspace{-4mm}
    \label{fig:enter-label2}
\end{figure}
\vspace{-4mm}
\subsection{Contrastive Pretraining}
\vspace{-1mm}
In the pretraining stage, we obtain intra-speaker clusters of speaker embeddings in a separate process similar to the experiments in Section \ref{clus_exp}, where the only difference is in the number of clusters $N$ since we don't have a prior about categories on emotion-unlabeled data. \mycomment{\textcolor{orange}{explain N}} 
A variant NT-Xent\cite{simclr} loss is used as an objective in the training:


\begin{equation}
\begin{gathered}
    l = -log\frac{exp(sim(z_i,z_j)/\tau)}{\sum_{k=1}^{N/2}exp(sim(z_i,z_k)/\tau)_{[k\ne i]}}
\end{gathered}    
\end{equation}
where $z_i,z_j$ is the positive pair and $z_i,z_k$ are the negative pairs for a given utterance. The similarity function $sim(x,y)=x^Ty/||x||.||y||$ calculates the cosine similarity and $\tau$ denotes the temperature parameter.  

\textbf{Soft-sampling:} For each utterance, we select one positive and $N/2$ negative utterances based on intra-speaker cluster labels. Due to rough clustering, when sampling the negative examples, we employ a soft-sampling strategy, selecting one negative sample from each of the $N/2$ intra-speaker clusters that are farthest from the positive cluster center. The model architecture consists of an encoder followed by a contrastive learning head, as shown in Fig.2(a) and Fig. 3.
\vspace{-4mm}
\subsection{Contrastive Pretraining for Multi-Task Design}
\vspace{-1mm}

\begin{figure}[t]
    \centering
    \scalebox{0.5}{\includegraphics{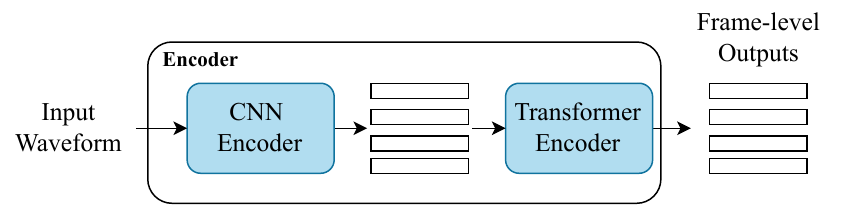}}
    \vspace{-3mm}
    \caption{The encoder architecture utilized in the networks.}
    \vspace{-4mm}
    \label{fig:MTT}
\end{figure}

Given the success of the transfer learning from speaker recognition to SER due to their connection, we also propose a multi-task learning (MTL) strategy to utilize available speaker labels. The proposed multi-task framework includes shared encoder layers along with two separate heads: contrastive learning and speaker classification head which can be seen in Figure 2(b). The contrastive learning head is trained with the proposed objective in Section 3.1; while the speaker classification head is trained with the cross-entropy loss with speaker labels. Along with the multi-task framework, speaker adversarial setting is also experimented, by including a Gradient Reversal Layer (GRL) just before the speaker classification head.


\vspace{-4mm}
\subsection{Speech Emotion Recognition}
\vspace{-1mm}
After pretraining on a large-scale, emotion-unlabeled dataset, the model is trained in a supervised manner on a smaller dataset with categorical emotion labels. During supervised training, we introduce a freshly initialized classification head on top of pre-trained encoder layers. This classification head comprises an average pooling layer, a dense projection layer with rectified linear unit (ReLU) activation, and a dense output layer with softmax activation. In this stage, we fine-tune the pre-trained layers in conjunction with the classification head, utilizing cross-entropy loss and emotion labels. The diagram can be seen in the Figure \ref{fig:enter-label2}(a).


\label{sec:typestyle}


\vspace{-4mm}
\section{Experiments}
\vspace{-2mm}

In this section, we report the effect of our proposed pretraining strategies with only contrastive loss and multi-task learning on SER performance when dealing with a limited amount of labeled data. We have evaluated our strategies independently and in conjunction with wav2vec2.0 to clearly discern their effect on emotion recognition performance.

\label{sec:typestyle}
\vspace{-4mm}
\subsection{Experimental setup}
\vspace{-1mm}

\textbf{Datasets:} During pretraining, we utilize voxceleb2 \cite{Chung2018VoxCeleb2DS} as an emotion-unlabeled dataset, known for its diverse emotional contexts \cite{Williams2019DisentanglingSF}, aligning with our intra-speaker clustering approach. In supervised SER training, we separately employ two labeled emotion datasets, IEMOCAP and CREMA-D. We focus exclusively on \textit{Anger, Happiness, Neutral, Sadness}, establishing a speaker-independent emotion recognition scenario. For the IEMOCAP corpus, we only use improvised utterances and create 5-fold training and test splits following the leave-one-session-out rule described in \cite{Morais2022SpeechER} and \cite{Pappagari2020XVectorsME},  excluding a small subset from one of the test speakers for validation. For CREMA-D, we use training data from 64 speakers, with 8 for validation and 19 for testing.


\textbf{Baselines}: In our basic SER experiments, we establish three baselines: \textit{No-pretraining}, which involves initializing the model randomly before supervised SER training, without any pretraining; \textit{No-pretraining (small)}, which has a smaller architecture with only 2 transformer layers to assess the impact of overfitting; and \textit{Pretraining w/ spk classification} which employs pretraining the model with encoder followed by only speaker classification head and loss, similar to the methodology in \cite{Pappagari2020XVectorsME}.
For SER experiments based on wav2vec2.0, we utilize a smaller version of the original \textit{wav2vec2.0} as the baseline pretraining method.



\textbf{Model Architecture \& Training}: In our pretraining and basic SER experiments, our proposed methods and baseline models, have the same encoder architecture, which is based on wav2vec2.0 \cite{wav2vec2}. This encoder architecture includes a feature extractor and a transformer encoder, similar to wav2vec2.0, but with a more compact design featuring only 6 transformer layers. The contrastive learning head includes an average pooling layer for frame-level outputs, followed by two dense layers featuring ReLU and tanh activation functions shown in Figure \ref{fig:enter-label2}, respectively. The speaker and emotion classification heads have a similar structure as the contrastive head but use softmax activation at the output layer, shown in Figure \ref{fig:enter-label2}. In the speaker adversarial setting, we introduce an additional GRL layer after pooling and before the speaker classification head. All the proposed models take the raw waveform of an utterance as input. 

During pretraining, we segment the input utterances into 4-second intervals and perform offline intra-speaker clustering with $N=20$. The models are pretrained for 250k steps using the AdamW optimizer with a batch size of 8. In the supervised SER training that follows, the model undergoes 30 epochs of training with a learning rate of 1e-5, stopping based on the validation accuracy. We repeat each supervised SER training 5 times with different initialization seeds and measure unweighted average recall (UAR) during the evaluation. For the SER experiments based on wav2vec2.0 reported in Table \ref{tab:ser_wav2vec}, the baseline \textit{wav2vec2.0}\footnote{https://github.com/facebookresearch/fairseq/blob/main/examples/wav2vec} with 6 transformer layers, is pretrained for 400k steps on voxceleb2. We then fine-tune this model with our strategies on voxceleb2 for an extra 50k steps. The feature extractor and transformer layers of fine-tuned wav2vec2.0 are utilized in the supervised SER training.

\vspace{-4mm}
\subsection{Results and Discussion}
\vspace{-1mm}
According to the results in Table \ref{tab:ser}, our proposed contrastive strategy, denoted as \textit{Pretraining w/ proposed contrastive}, demonstrate a significant improvement in SER compared to cases with no pretraining in both datasets. We note that pretraining with supervised speaker classification also leads to substantial improvements in both datasets, consistent with findings in \cite{Pappagari2020XVectorsME}. The proposed multi-task learning approach, denoted as \textit{Pretraining w/ proposed MTL}, leverages the inherent connection between speaker and emotion recognition, while simultaneously considering intra-speaker variations and obtains the best performance in the IEMOCAP corpus.  We observe that speaker adversarial network degrades performance, indicating that  trying to remove speaker information has a negative impact and supports the connection between speaker and emotion recognition. In CREMA-D, the speaker classification baseline performs exceptionally well, possibly due to the presence of normalized linguistic content, creating ideal conditions for discriminating emotions through speaker embeddings, as discussed in Section 2. Overall, these results underscore the effectiveness of our multi-task learning method and highlight the strong relationship between emotion and speaker recognition.

\begin{table}[!t]

\centering
\caption{SER results, in terms of mean UAR.}
\label{tab:ser}
\vspace{-3mm}
\scalebox{0.80}{\begin{tabular}{c|cc}

\hline
\textbf{Pre-Training}                        & \textbf{\begin{tabular}[c]{@{}c@{}}IEMOCAP \\ (UAR) \end{tabular}} & \textbf{\begin{tabular}[c]{@{}c@{}}CREMA- D \\ (UAR)\end{tabular}} \\ \hline 
\textit{No pretraining (small) }                                   & \textit{58.37 \mycomment{($\pm$ 0.7)}}                                                           & \textit{65.12 \mycomment{($\pm$ 0.4)}}                                                    \\  
\textit{No pretraining}                                   & \textit{58.54 \mycomment{($\pm$ 0.6)}}                                                           & \textit{64.15 \mycomment{($\pm$ 0.3)}}                                                    \\  
\textit{Pretraining w/ spk classification} & \textit{67.57 \mycomment{($\pm$ 1.0)}}                                                          & \textbf{\textit{75.23}}                                                             \\  
Pretraining w/ proposed contrastive                   & 65.50 \mycomment{($\pm$ 0.9)}                                                                   & 70.44                                                             \\ 
Pretraining w/ proposed  spk ADV   & 64.48 \mycomment{($\pm$ 0.9)}                                                                    & 66.58                                                             \\ 
\textbf{Pretraining w/ proposed  MTL}         & \textbf{69.16  \mycomment{($\pm$ 0.8)} }                                                                  & 73.80                                                            \\ \hline
\end{tabular}}

\end{table}

\begin{table}[t]
\vspace{-2mm}

\caption{SER results with wav2vec2.0, mean UAR.}
\label{tab:ser_wav2vec}
\vspace{-3mm}
\begin{adjustbox}{width=0.9\columnwidth,center}
\begin{tabular}{c|cc}
\hline
\textbf{Pre-Training}                        & \textbf{\begin{tabular}[c]{@{}c@{}}IEMOCAP \\ (UAR) \end{tabular}} & \textbf{\begin{tabular}[c]{@{}c@{}}CREMA- D \\ (UAR)\end{tabular}} \\ \hline
\textit{wav2vec2.0 \cite{wav2vec2}}                          & \textit{72.14 \mycomment{($\pm$ 1.4)}}                                                                  & \textit{80.78 \mycomment{($\pm$ 0.5)}}                                                    \\  
FT wav2vec2.0 w/ proposed contrastive &  72.78  \mycomment{($\pm$ 1.2)}                                                        & 81.72                                                             \\ 
\textbf{FT wav2vec2.0 w/ proposed MTL} & \textbf{73.80 \mycomment{($\pm$ 1.2)}}                                                                    &  \textbf{83.01}                                                             \\ \hline
\end{tabular}
\end{adjustbox}
\vspace{-4mm}
\end{table}

In Table \ref{tab:ser_wav2vec}, baseline \textit{wav2vec2.0} model, pretrained with voxceleb2, performs impressively well as a pretraining method for SER, underscoring its effectiveness. Fine-tuning this baseline with our contrastive learning strategy, \textit{FT wav2vec2.0 w/ proposed contrastive}, seems leading to minor improvements in both datasets. Fine-tuning wav2vec2.0 with our proposed multi-task setting,  \textit{FT wav2vec2.0 w/ proposed MTL}, yields substantial enhancement, highlighting the effectiveness of our approach.

\vspace{-5mm}
\section{Conclusion}
\vspace{-3mm}
\label{sec:typestyle}
Our research reveals the potential of speaker embeddings for enhancing SER task, even with limited labeled data. Our study establishes a direct link between emotions and state-of-the-art speaker embeddings through intra-speaker clusters. Our novel contrastive pretraining approach on emotion-unlabeled datasets, based on these clusters, significantly improves SER performance, whether used alone or in multi-task settings. Our findings not only advance our understanding of speaker embeddings and emotions but also provide practical solutions for data scarcity in SER. As a future work, we intend to extend the analysis of emotion information in speaker embeddings, analyzing other factors which potentially affect the appearance of that information.



\footnotesize
\bibliographystyle{IEEEbib}
\bibliography{strings,refs}

\end{document}